# Bans vs. Warning Labels: Examining Bystanders' Support for Community-wide Moderation Interventions

Shagun Jhaver

Rutgers University, shagun.jhaver@rutgers.edu

Social media platforms like Facebook and Reddit host thousands of user-governed online communities. These platforms sanction communities that frequently violate platform policies; however, public perceptions of such sanctions remain unclear. In a pre-registered survey conducted in the US, I explore bystander perceptions of content moderation for communities that frequently feature hate speech, violent content, and sexually explicit content. Two community-wide moderation interventions are tested: (1) community bans, where all community posts are removed, and (2) community warning labels, where an interstitial warning label precedes access. I examine how third-person effects and support for free speech influence user approval of these interventions on any platform. My regression analyses show that presumed effects on others are a significant predictor of backing for both interventions, while free speech beliefs significantly influence participants' inclination for using warning labels. Analyzing the open-ended responses, I find that community-wide bans are often perceived as too coarse, and users instead value sanctions in proportion to the severity and type of infractions. I report on concerns that norm-violating communities could reinforce inappropriate behaviors and show how users' choice of sanctions is influenced by their perceived effectiveness. I discuss the implications of these results for HCI research on online harms and content moderation.

**ACM Reference Format:**

Shagun Jhaver. 2024. Bans vs. Warning Labels: Examining Support for Community-wide Moderation Interventions: ACM Transactions of Computer-Human Interaction. ACM, New York, NY, USA, 24 pages. NOTE: This block will be automatically generated when manuscripts are processed after acceptance.

## 1 INTRODUCTION

Today, social media companies like Facebook, Reddit, and YouTube are subject to increasing pressure from law enforcement agencies and private actors worldwide to act more forcefully to address online harm. In response, platforms are investing in better monitoring of content posted on their sites and experimenting with a new range of moderation strategies. One such strategy is to sanction an entire online community that hosts inappropriate content to quickly tamp down its spread. Such actions are prevalent since they let platforms demonstrate their commitment to reducing harm. However, these sanctions can be deemed excessively broad. By removing non-offending speech alongside offending content, they raise questions about violating some users' freedom of speech.

I examine the public response to hypothetical instances of community-wide moderation interventions in the context of three categories of norm-violating online communities: those frequently featuring hate speech, violent content, and sexually explicit content. Informed by the third-person effects literature [23, 43], I evaluate the role that perceived influences of communities hosting such content have on users' willingness to support two types of community-wide sanctions prevalent across social media:

1) *Community bans*, where all community content and access to it are permanently removed [15, 92]. Sites like Facebook (on Facebook Groups) [132], Reddit (on its subreddits) [15], Discord (on its servers) [20], and YouTube (on its channels) [94] frequently deploy such bans.

2) *Community warning labels*, where a warning label precedes access to community content. Quarantines on Reddit [12, 110] exemplify this intervention. Similarly, Facebook shows a warning message to users about to join a group that allows posts violating the site's community standards [82]. Other examples of warnings preceding content access have been observed on Instagram, search engines, and Pinterest [14].

My analysis focuses on understanding *bystanders*' perceptions of these sanctions. Here, bystanders refer to users who are not already actively contributing to an online community but encounter it during their regular social media use. When assessing platform-enacted sanctions, bystanders constitute an essential stakeholder even though they are not the ones directly impacted. As I argue in more detail in the next section, fairness perceptions of bystanders can shape their trust in the platforms, their perceived social norms, and how they behave on the site.

This article also explores how *free speech support* influences user perceptions of community-wide interventions on any social media site. In the United States, the First Amendment protects speech from government censorship. This protection applies to federal, state, and local government actors, including courts, police officers, public schools, and universities. Technically, the legal protections of freedom of speech do not apply to private parties like social media platforms [39]. However, since a few large platforms like Facebook and Reddit have become the central avenues for our public discourse, they essentially serve as public spaces. Platforms' moderation policies and decisions regarding content sanctions distinctly shape how we participate in those spaces [117]. Therefore, when companies make major moderation decisions, such as community-wide bans or sanctions of public figures [52, 98], they raise issues of free speech and censorship that warrant careful consideration.

I evaluate third-person effects and free speech support as influencing factors because, as I detail in the next section, prior literature (e.g., [58, 96]) has established their relevance in shaping user attitudes about content moderation. Interrogating these issues, I present results from a nationally representative survey of 1,023 US adults. During this survey, I asked participants how they would like platforms to handle any community that frequently hosts inappropriate content. To be clear, this was not a field experiment on a specific site; I grounded this survey questionnaire within the context of all social media platforms that host online communities or groups. By asking participants about hypothetical communities that they know relatively little about, I simulate how bystanders (as opposed to already active contributors) would react to inappropriate communities.

I found that about 90% of participants opted to select a platform-enacted moderation action (a ban or a warning label) to regulate each of the three categories (i.e., hate speech, violent content, sexually explicit content) of norm-violating online communities. My regression analysis of this survey data finds that users' perceptions of norm-violating communities' negative effects on others significantly influence their support for both community bans and community warning labels. Further, support for free speech significantly influences users' support for warning labels that precede community access. My qualitative analysis of open-ended responses from this survey aligns with these insights and reveals additional perspectives that shape respondents' choice of sanction. I found a widespread concern that norm-violating communities could influence susceptible users, especially children, to develop more extremist views and even engage in violent acts. Additionally, many participants desire a greater nuance in the enactment of community-wide moderation sanctions. The expected empirical effects of different moderation actions also influenced some participants' choices.

A single community-wide sanction decision can influence or prohibit all subsequent community activity, thereby regulating thousands of bad actors at once. Therefore, such sanctions present an enticing strategy for platforms to address the ever-growing challenges of scale. However, their deployment also raises, perhaps more strongly than ever, the longstanding, familiar tensions of striking an appropriate balance between reducing content-based harms and ensuring freedom of expression [122]. Though prior work *theoretically* examined the tensions involved in community-wide sanctions [12], this article evaluates *broader public opinions* on these issues. In doing so, it adds to a growing body of HCI and social computing research [14, 15, 72, 107, 130] on examining the effectiveness of different moderation actions. This paper also contributes to our understanding of how the adult US population perceives distinct categories of online harms and the tradeoffs involved in various approaches to address them. Recognizing the levels of public support for different interventions can help guide future community-wide moderation practices on social media platforms.

## 2 LITERATURE REVIEW AND HYPOTHESES

### 2.1 Content-based Harms

The study I conduct here focuses on the harms caused by viewing inappropriate content on social media platforms [55]. Specifically, it explores three types of content-based harms examined in prior literature [58]: hate speech, violent content, and sexually explicit content. Prior research on addressing hate speech has focused on the challenges of defining and bounding it [70] and developing natural language processing tools to detect it [124]. Violent content includes a wide range of materials, such as images and videos of graphic violence, police brutality, self-harm, and eating disorders [13, 24, 104]. Díaz and Hecht-Felella [25] have shown that social media guidelines against violent content are usually opaque and grant platforms immense discretion in how they choose to enforce them. Platforms' attempts to moderate sexually explicit content often end up censoring sexual health education [10] or disproportionately targeting sex workers, queer individuals, and even sexual assault survivors [3, 95, 112]. Yet sexually explicit content could also include non-consensual pornography and sexual exploitation of minors, which could damage vulnerable individuals. The current study examines approaches to addressing these harms when they frequently occur in an online community.

Previous studies have found that exposures to disturbing posts online are associated with PTSD and depressive symptoms [121] and can induce self-harming behaviors [44]. Online platforms often develop ad hoc content moderation approaches to deal with such harmful content [89]. Community-wide bans and warning labels are examples of how platforms attempt to address harm. However, we do not yet fully understand how end-users perceive such moderation interventions. Given the especially broad-brush impact of such actions, it is vital to examine bystanders' perceptions of them as measures to address content-based harm. Therefore, this inquiry forms the focus of my research.

### 2.2 Content Moderation and Community-wide Moderation Interventions

As platforms grow, the challenge of efficiently and economically moderating high volumes of user-generated content becomes increasingly salient [36, 71]. To address this challenge, each platform has created its own set of complex, human-machine collaborative systems [51] that apply moderation actions at different levels of granularity – from removing a post or de-platforming a user [52] to sanctioning an entire community [12, 14].

My focus in this paper is on community-wide moderation sanctions. I distinguish these from *user-level* sanctions that apply to only a single post or poster. Prior research examined user perceptions of the censorship of hateful messages [43, 49, 96] and user reactions to experiencing sanctions themselves [50, 126]. However, as far as I know, third-party user perceptions of community-wide moderation interventions have yet to be explored. This article seeks to take the first steps toward filling this gap because user support and attitudes toward content moderation should be critical considerations for platforms when designing their moderation policies [96], especially regarding sweeping actions that impact entire communities.

I specifically address communities like Facebook Groups, Discord servers, and Reddit's subreddits, which constitute the *middle levels* of governance – they serve the end-users at the bottom level but are themselves accountable to platform administrators at the topmost level [56]. These communities are governed by volunteer moderators, usually selected from users active within the community [51, 73, 74]. These moderators work largely autonomously, creating their community submission guidelines, reviewing all posted content, and meting out sanctions when warranted [108]. However, in some cases, when a community is seen to be repeatedly violating the platform's terms of service with impunity, the platform staff exercise their authority to regulate it.

Prior HCI research computationally evaluated the effects of these interventions on the behavior of affected community members, finding their utility in decreasing hate speech usage and radicalization [14, 15, 47, 120]. I add to this literature by surfacing the perspectives of *bystanders*, a large portion of users who are not directly impacted by the community-wide moderation but are, nonetheless, essential stakeholders for platforms. When bystanders perceive moderation of other users' content as fair, they are more likely to reduce inappropriate behavior through social learning [7] and increase their efforts to enforce norms, e.g., through reporting infractions [84]. Prior content moderation research centered on bystanders has analyzed how information visibility around the identity of moderators [9] and rationale for moderation decisions [57] influence bystanders. I add to this research by examining the factors that play a role in bystander support for bans and warning labels on inappropriate online communities.

### 2.2.1 Banning Versus Warning

Singhal et al. [111] classified platforms' approaches to enforce their policies in two categories: *hard* and *soft* moderation. Hard moderation involves removing problematic content or entities from platforms, whereas soft moderation does not remove content but adds a warning label before it to inform users about potential issues. Hard moderation promotes public safety but can often be viewed as contradicting principles of open participation, digital equity, and freedom of speech [33, 119, 123]. HCI researchers have also highlighted the ways in which discriminatory moderation practices disproportionately remove content submitted by Black users, LGBTQ+ users, sex workers, sexual educators, and activists [3, 22, 45, 95]. On the other hand, soft moderation, which has largely been deployed and studied in the context of combating misinformation [79], limits users' ability to interact with questionable content [60, 78, 101, 109]. However, it leads to the implied truth effect, where unlabeled content is perceived as more credible [91]. In the context of community-wide moderation, bans and warning labels can be seen as instances of hard and soft moderation, respectively.

Decisions about whether a platform should ban an offensive community or merely put a warning label on it can be complex and subjective. Platform policies usually address the moderation of individual pieces of content. However, it can be difficult for platforms to determine and specify which behavioral features of an entire community warrant banning it instead of just informing users about offensive content through labeling.

While research on both community bans [15] and community warning labels [12, 14, 110, 122] continues to expand, I have found no prior study that directly compares user preferences for the two interventions. However, in the related context of user preferences for moderating individual news articles, user survey data from Atreja et al. [5] show that attaching warning labels is a more popular option than removing the news articles containing misinformation. Similarly, Wihbey et al. [127] found that across four democracies, people prefer that platforms deploy content labeling rather than shutting down accounts to address misinformation and hate speech. These insights suggest that putting a warning label before a community hosting inappropriate content would be more popular than banning it. On the other hand, if an inappropriate community is allowed to remain online, it can continue to host many norm-violating posts on an ongoing basis. This concern may motivate bystanders to prefer bans over warning labels. Understanding which of these moderation sanctions is more popular can offer policy guidance to platform administrators for regulating online communities [56]. To verify whether prior findings on the popularity of warning labels over bans for user-level sanctions [5, 127] also apply in the context of community-wide moderation, I raise the following hypothesis:

**H1:** For each type of norm-violating community, the average support for adding warning labels before it will exceed the average support for banning it.

### 2.3 Third-Person Effects

Davison [23] originally proposed the third-person effect (TPE) hypothesis, which posits that individuals exposed to a mass media message will expect that the communication's greatest impact "will not be on 'me' or 'you,' but on 'them' – the third persons." According to Davison, TPE has two distinctive components: (1) *the perceptual component*, which predicts that people perceive themselves as invulnerable to the effects of media and perceive others as more greatly affected, and (2) *the behavioral component*, which predicts that the perceptions of negative effects on others would lead to certain cognitive, attitudinal, and behavioral consequences [16, 90].

The perceptual component of TPE is widely established across a wide range of message types, effect domains, and populations. Historically, TPE was first observed in perceptions of the effects of traditional mass media, such as television violence [103], defamatory news articles [40], product advertising [42], and rap music [77]. However, Peiser and Peter [90] argued that third-person perceptions extend beyond media effects and reflect a general tendency of individuals to underrate others' education. Further, Flanagin and Metzger [32] pointed out that the Internet is used in a manner similar to other mass media, such as newspapers, books, and television, for information-retrieval and information-giving purposes. This provided the basis for extending the study of TPE to the use of blogs and social media [64, 66, 106].

Researchers subsequently broadened TPE analysis to study activities dissimilar to passive message receipts reminiscent of traditional media, such as interactive communications [67]. For example, Paradise and Sullivan [87] surveyed 357 undergraduates in the Northeastern US and found that young people perceived Facebook usage to have a more significant negative impact on others' personal relationships, future employment, and privacy than on their own. Examining the influence of online advertising, Lim [68] found that the self-other disparity in TPE is particularly notable when the content of persuasive communication is socially undesirable or harmful.

Extending this prior literature and applying TPE to the context of inappropriate online communities on social media, I expect that users would perceive such communities to exert greater influence on others than on themselves. The second hypothesis can thus be raised:

**H2:** For each type of norm-violating community, participants will perceive a greater effect of that community on others than on themselves.

*2.3.1 TPE and Community Bans*

The behavioral component of TPE predicts that perceiving others as more influenced than oneself will lead individuals to take remedial actions, such as supporting censorship of media content [23, 66]. Scholars have argued that while the perceptual component of TPE is intriguing, the behavioral component is more relevant to social researchers and media practitioners [77, 87]. For example, Gunther [41] contends that TPE's significance lies in predicting individuals' readiness to advocate action to shield others from perceived negative media impact.

Most prior literature on TPE consequences has found that it leads to support for censorship or governmental regulation of the media [37, 41]. In particular, researchers have studied how socially undesirable media messages, such as political advertisements against liked candidates, result in support for regulation [18]. Examining the previously proposed theoretical rationales for how third-person effects shape perceptions about censorship, Chung and Moon [17] concluded that the media's presumed effect on others (PME3) is a stronger predictor of censorship attitudes than the other-self disparity in perceived media effects. In line with this, Riedl et al. [96] found that the perceived effects of social media content on others are a significant predictor of support for content moderation.

Accordingly, I expect that in encountering inappropriate online communities, the perceived effect on others (PME3) will likely result in support for community-wide bans. Therefore, I raise the third hypothesis regarding the TPE behavioral component:

**H3:** For each type of norm-violating community, its perceived effects on others will be positively related to support for the platform's banning of that community.

By testing this hypothesis, I seek to clarify whether concerns about others influence users' moderation preferences for inappropriate communities, similar to how they affect moderation preferences for inappropriate posts.

*2.3.2 TPE and Community Warning Labels*

While support for censorship as a behavioral component of TPE continues to be the mainstay of third-person effects researchers, a growing body of studies has also identified corrective actions as another behavioral outcome of TPE [69, 116]. In contrast to restrictive actions, such as censorship or government regulation, corrective actions "refer to individuals' engagement in reactive action against potentially harmful influence." [68]

In the context of online content moderation, researchers have proposed calls for media literacy interventions as one of the most prominent corrective efforts [49, 63]. Jang and Kim [49] found that people with a greater third-person perception were more likely to support the media literacy approaches to address fake news. In line with this, perceived harms of an offensive community on others are likely to result in support for improving media literacy about that community. Putting warning labels on inappropriate communities is a potent media literacy intervention since it aims to educate users on critically evaluating the community's content. Therefore, I examine the following hypothesis:

**H4:** For each type of norm-violating community, its perceived effects on others will be positively related to support for the platform's adding a warning label on that community.

## 2.4 Support for Freedom of Speech

The term "free speech" refers to both a philosophical stance and a legal doctrine. As a philosophical stance, it indicates "a fundamental moral requirement that agents be free to express themselves and communicate with others." [48] As a legal doctrine, it refers to the legal right to freedom of speech, which is constitutionally protected as a human right in most democracies [88]. Historically, philosophers have considerably influenced lawyers and judges responsible for the constitutional doctrines of freedom of speech [1]. Indeed, many political theorists assume that the legal doctrine of free speech is simply the legal codification of the moral right to freedom of speech [48]. In this context, free speech experts broadly construe "speech" to include not just (spoken or penned) words but the whole range of expressive acts produced to communicate ideas, including paintings, photographs, and performances [1, 11, 48].

Scholars have long debated over the extent to which the free speech principle should protect against the regulation of different types of harmful speech, such as targeted vilification and advocacy for exclusionary policies [131]. Those who support deplatforming individuals or groups holding hateful views are often criticized as being anti-free speech [38, 52]. They defend themselves by arguing that hate speech creates a hostile environment that silences marginalized groups, and thus, denying hate speech is necessary to protect the free speech rights of precarious members of society [54, 58]. Analyzing the historical ideas that have shaped the US free speech laws, Bejan [8] highlights a fundamental conflict between two free speech concepts from ancient Athens that underlie our current controversies: *isegoria*, which describes "the equal right of citizens to participate in public debate" and *parrhesia*, which refers to "the license to say what one pleased, how and when one pleased, and to whom." The First Amendment brings these concepts together, but the tensions between them help explain the frustrating shape of contemporary debates [8].

Although Section 230 of the Communications Decency Act immunizes social media platforms from First Amendment scrutiny [26, 34], the US Supreme Court recognizes these platforms as "probably the most powerful free speech vehicle available to citizens." [28] It has been widely argued that given platforms' increased importance for public and political discourse in recent years, they should be required to uphold free speech principles [6, 28, 35, 62]. Allegations about platforms' abuse of power, political bias, viewpoint discrimination, and lack of due process protections have led to calls for regulation that prioritizes the free speech interests of members of the public over the free speech interests of platforms [85, 97].

While both "censorship" and "content moderation" refer to actions taken by a governing entity to delete expressive content, they have different connotations [62, 113]. Whether any given governance intervention is condemned as censorship or accepted as content moderation depends on the subjective perspective of observers and the source of intervention within the institutional hierarchy [20]. For example, state-led interventions, which could ban entire websites, domains, or protocols [61], are more likely to be seen as censorship. In contrast, regulation efforts by moderators of well-run, small online communities [51, 74, 76] are more likely to be viewed as content moderation. However, in recent years, critics and lawmakers complained that content moderation efforts of platforms like Facebook, YouTube, and Reddit often amount to censorship [65, 113].

Prior work suggests that attitudes toward censorship and online moderation are drawn along partisan lines, but it remains unclear which divergent underlying factors account for this partisan gap [2]. To examine how people perceive such tensions [59], specifically in the context of community-wide moderation interventions on social media, it is crucial to examine how individuals' support for freedom of speech impacts their moderation attitudes.

### 2.4.1 Support for Free Speech and Community Bans

Prior research has explored connections between attitudes toward free speech and attitudes about content moderation, but findings are largely mixed. While Americans generally support free speech, their tolerance for hate speech is low [115]. Support for free speech does not necessarily indicate opposition to moderation [58, 83]. For example, Riedl et al. [96] found that support for free speech did not significantly impact attitudes about moderation. However, a related measure, opposition to censorship, significantly negatively affected support for social media content review. Jang and Kim [49] argued that support for free speech reduces support for regulating fake news despite the existence of TPE. Guo and Johnson [43] conducted a survey experiment with 368 US university students, showing that respondents' support for freedom of expression significantly negatively predicted their support for Facebook's content moderation of anti-LGBT speech but not for the moderation of racist or sexist speech. Taken together, these insights suggest that individuals' appreciation of the expression rights in the abstract may give only an incomplete picture of their tolerance for opposing expressions [83]. Since it is unclear how support for free speech might impact users' preferences for community-wide moderation, I ask the following research question:

**RQ1:** For each type of norm-violating community, how does support for freedom of expression relate to support for the platform's banning of that community?

*2.4.2 Support for Free Speech and Community Warning Labels*

Generally, support for free speech would be expected to reduce support for *any* restrictive actions, including the placement of warning labels before communities. However, in contrast to community bans, adding a warning label to an online community does not remove its content and lets users decide whether to engage. I expect this intervention to be perceived as free speech preserving, especially because it contrasts with community bans that let platforms unilaterally remove all prior community posts and stall any further activity. Prior research by Jhaver and Zhang [58] also showed that free speech supporters prefer an approach to content curation that emphasizes individual choice instead of top-down content removal by social media platforms. Therefore, I raise the following hypothesis:

**H5:** For each type of norm-violating community, participants' support for freedom of expression will be positively related to support for the platform's adding a warning label on that community.

Free speech is a fundamental American value that relates to the tensions between censorship and safety online. In the current political climate, platforms have increased their self-proclamations as free-speech defenders, yet they continue to make controversial moderation decisions [30, 31]. Given this context, **RQ1** and **H5** seek to examine the extent to which free speech values currently influence users' perspectives on platform-enacted community-wide moderation.

## 3 METHODS

Rutgers University's IRB examined this study and determined it to be exempt from a full review on Dec 30, 2022. My inclusion criterion for the survey participants was all adult internet users in the US. I recruited participants through Lucid Theorem,[1] an academically-oriented survey platform that compares favorably to other alternatives like Amazon's Mechanical Turk [21]. Lucid employs quote sampling procedures to gather responses from its large online panel of potential participants. It provides researchers access to samples that target representativeness to the US population across several demographic categories, including gender, race/ethnicity, level of education, age, and political affiliation. On average, our participants took four minutes to complete the survey. Each participant received a compensation of $1.5 for their participation. This study was preregistered at OSF.[2]

To begin with, I designed an initial draft of my survey questionnaire to operationalize the measures relevant to my hypotheses. I drew from prior survey research testing similar measures to inform this operationalization. Once I had an initial draft of my survey questions, I took several steps to improve the questionnaire quality. First, I reached out to students and colleagues at my institution who were not involved with the project to get their feedback. Nine such individuals agreed to assist me with this survey pretest. I set up my initially drafted survey questions in Qualtrics and observed these testers as they completed the survey. I asked them to speak aloud about their perspectives, confusions, and challenges with the questions and sought suggestions on how I could improve the wording of the questions. Following these tests, I incorporated the feedback I received therein to revise my survey questions. I ensured that the survey interface worked well for both Desktop and mobile users.

---

[1] https://lucidtheorem.com

[2] https://osf.io/pnv29/?view_only=083d7f5dad1a46a1a32321bf2313a7e4

Next, I piloted the survey with a sample of 30 Lucid participants. I inserted frequent prompts for open-ended feedback on the questionnaire quality. Since this pilot drew respondents from the same pool as the main survey, its feedback helped ensure that the survey questions did not have any critical shortcomings that my academic testers in the previous testing phase could have missed. After this pilot, I conducted another round of iteration over my survey questionnaire to reach the desired quality. Next, I rolled out the survey and collected data, preprocessed it, and built hierarchical regression models [29] to examine it. I describe below in more detail how I used survey items to measure each relevant variable along with the variable's mean (M), standard deviation (SD), and Cronbach's alpha ($\alpha$) values from the processed survey data. I also note the socio-demographic variables I controlled for in these models.

The survey began with a consent form informing respondents that the data retrieved from them would remain anonymized and stored on a secure server. Next, they were shown a page entitled "What is an Online Community." This page explained what an online community means and showed a screenshot of r/AskReddit home page as a typical example, although it clarified that this study concerns all social media platforms (and not Reddit specifically). It also noted that a few online communities frequently host inappropriate content such as hate speech, violent posts, or sexually explicit posts and informed respondents that the survey would ask their perspectives on how they would like platforms to handle such communities. Next, participants went through three blocks of similar questions about communities featuring hate speech, violent content, and sexually explicit content. The order in which these three blocks appeared to each participant was randomized. Each such block began with indicating the norm-violating category that the following questions related to and defined that category. I used the following definitions, adopted from [58]:

- Hate speech: *"Hate speech includes speech that is dehumanizing, stereotyping, or insulting based on identity markers such as race/ethnicity, gender, sexual orientation, religion, etc."*
- Violent content: *"Violent content includes threats to commit violence, glorifying violence or celebrating suffering, depictions of violence that are gratuitous or gory, and animal abuse."*
- Sexually explicit content: *"Sexually explicit content includes content showing sexual activity, offering or requesting sexual activity, female nipples (except breastfeeding, health, and acts of protest), nudity showing genitals, and sexually explicit language."*

These definitions were originally inspired by the language used by the Facebook site when reporting any post under the category of hate speech, violence, and nudity, respectively. Note that in selecting these content categories, I do not intend to pass any value judgement on them. Certainly, there are many scenarios where content classified under these categories could have prosocial outcomes. For example, graphic images of war violence can raise public awareness about the horrors of warfare [35, 118]. Similarly, sexually explicit content can spark movements and trends that promote body positivity and normalize different sexual practices [4, 19, 86]. Different cultures and societies would draw the line between the appropriate and the illicit differently [35]. Yet, most platforms currently seek to censor these three content categories in their policies and justify their content removals based on those policies [50, 102]. I, therefore, examine how users perceive the sanctions of communities frequently featuring them.

After defining the norm-violating content, each block asked questions (described in detail below) about how such communities would influence them and other people and how platforms should regulate them. By asking participants how they would react to these hypothetical communities, I simulated bystanders' perceptions about them.

Next, participants were asked questions about their support for free speech and how frequently they used social media recently. Finally, participants answered an open-ended question that asked them to reflect on their choices about the action platforms should take against communities containing different types of inappropriate content (i.e., hate speech, violent content, and sexually explicit content) and describe the reasoning behind those choices. This question also nudged participants, "If you chose differently for different types of content, please discuss why." Participants were reminded of the three available choices (i.e., community ban, community warning label, and neither) they had when they answered these prior questions. A brief description of these choices was also provided.

This survey was launched on May 13, 2023. **Table 1** presents the demographic details of my final sample of **1,023 participants** after data cleaning along the dimensions of gender, race/ethnicity, age, education levels, and Hispanic status. As **Table 1** shows, my sample closely mirrors the adult Internet population benchmarks from the US census along these demographic categories. Therefore, I expect my findings would generalize to adult US Internet users. Importantly, prior research has demonstrated that survey data from Lucid replicates well-studied experimental treatment effects and produces estimates of political attitudes and behaviors similar to those derived from probability samples of US adults [21, 80, 93].

**Table 1:** Demographics of survey participants.

|  | This study (%) | Adult US Internet population[a] (%) |
|---|---|---|
| **Gender** | | |
| Male | 48.0 | 48.6 |
| Female | 52.0 | 51.4 |
| **Race/Ethnicity** | | |
| White | 73.4 | 68.3 |
| Black | 12.9 | 9.3 |
| Other | 13.7 | 22.4 |
| **Age** | | |
| 18-29 | 21.2 | 17.4 |
| 30-49 | 38.0 | 29.5 |
| 50-64 | 23.3 | 25.6 |
| 65+ | 17.5 | 27.3 |
| **Education** | | |
| High school or less | 32.8 | 33.5 |
| Some college | 25.7 | 33.3 |
| College+ | 43.9 | 33.1 |
| **Hispanic** | | |
| Yes | 12.9 | 13.7 |

[a] Source: American Community Survey, 2021 [99]

### 3.1 Measures

*3.1.1 Perceived Effects of an Inappropriate Community on the Self and Others*

To measure the perceived effects of each norm-violating speech category – hate speech, violent content, and sexually explicit content – I asked survey participants to estimate the influence of engaging with communities containing it on the self and others. I adapted questions for each category based on survey instruments measuring third-person effects for that category in prior literature. Next, I present items measuring perceptions of each category on the self before I discuss how I modified these items to measure perceptions of effects on others.

For hate speech, participants rated the following two statements [43]:

1. Engaging with online communities that frequently contain hate speech would negatively influence my attitudes toward the targeted groups.
2. Engaging with online communities that frequently contain hate speech would negatively influence my attitudes toward anti-discrimination policies.

For violent content, participants rated the following statements [46]:

1. Engaging with online communities that frequently contain violent posts would lead me to view the world as a more dangerous place.
2. Engaging with online communities that frequently contain violent posts would lead me to think that aggression is acceptable.

For sexually explicit content, participants rated the following two statements [64, 135]:

1. Engaging with online communities that frequently contain sexually explicit posts would negatively influence my moral values about sex.
2. Engaging with online communities that frequently contain sexually explicit posts would negatively influence my attitudes toward the opposite sex.

For each question, the response levels ranged on a 7-point Likert-type scale from 1 (strongly disagree) to 7 (strongly agree). The two corresponding items for each speech category were averaged to create a measure of the perceived influence of communities featuring that category on the self (Hate speech: $M$=4.26, $SD$=1.89, $a$=.85; Violent speech: $M$=4.45, $SD$=1.61, $a$=.71; Sexually explicit speech: $M$=3.89, $SD$=1.83, $a$=.85).

For each category, I asked another two questions, replacing the words "me" with "other people" and "my" with "other people's." I averaged the responses to these questions to create an index of the perceived influence of communities featuring that category on others (Hate speech: $M$=5.00, $SD$=1.49, $a$=.90; Violent speech: $M$=5.09, $SD$=1.39, $a$=.75; Sexually explicit speech: $M$=4.52, $SD$=1.58, $a$=.90).

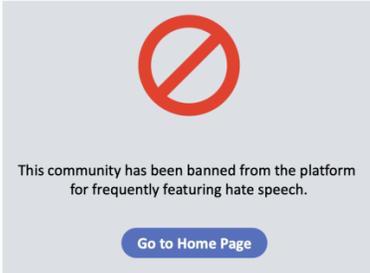

**Figure 1:** Survey question asking participants to rate their support for platforms banning communities that frequently contain hate speech.

### 3.1.2 Support for Freedom of Speech

This variable was assessed based on free speech measures examined in previous work [43, 58]. It included the following items: (a) In general, I support the First Amendment, (b) Freedom of expression is essential to democracy, (c) Democracy works best when citizens communicate in an unregulated marketplace of ideas, and (d) Even extreme viewpoints deserve to be voiced in society. I included the First Amendment statement[3] in the first question to clarify its meaning. Participants rated these four items on a Likert-type scale ranging from 1 (strongly disagree) to 7 (strongly agree). These four items were averaged to create an index for *"support for freedom of speech"* ($M$=5.39, $SD$=1.13, $\alpha$=.78). Note that these items measure free speech as both a philosophical stance and a legal doctrine. The high internal consistency of this scale ($\alpha$=.78) suggests that respondents perceive these two meanings of free speech as interconnected.

### 3.1.3 Support for Community-wide Moderation

*Support for the platform-enacted ban* of communities featuring each norm-violating speech category was assessed by asking participants to rate the following statement: "I support social media platforms banning any online community that frequently contains <speech category>." The responses for this statement ranged on a 7-point Likert-type scale, where 1="strongly disagree" and 7="strongly agree" (see **Figure 1**).

---

[3] The First Amendment to the United States Constitution states: "Congress shall make no law respecting an establishment of religion, or prohibiting the free exercise thereof; or abridging the freedom of speech, or of the press; or the right of the people peaceably to assemble, and to petition the Government for a redress of grievances."

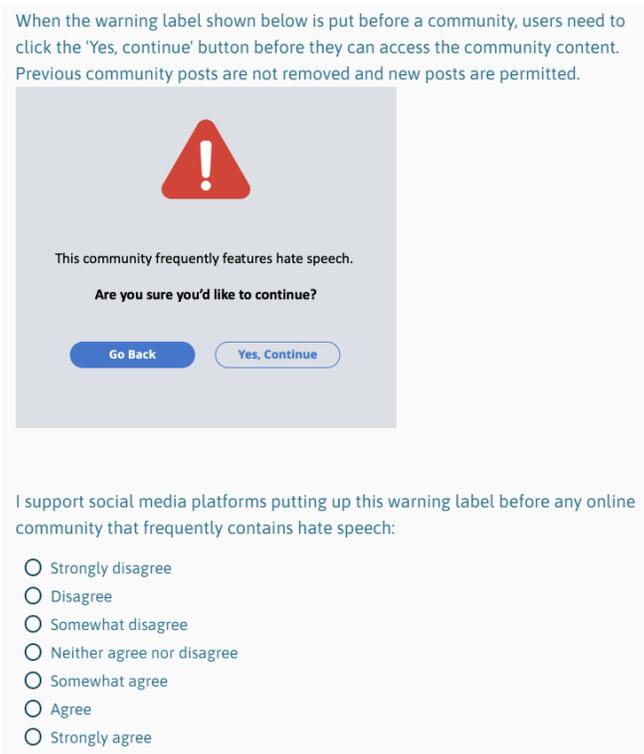

**Figure 2:** Survey question asking participants to rate their support for platforms adding a warning label before a community featuring hate speech.

To assess *support for adding a warning label* against communities for each category, I showed participants an example of a warning message preceding a community that lets users decide whether to continue accessing that community (see **Figure 2**). I asked participants to rate their support for inserting this community label on a Likert-type scale ranging from 1 (strongly disagree) to 7 (strongly agree).

Additionally, I assessed *choosing a community ban vs. a community warning label vs. neither* for each speech category. I asked respondents: *"Given a choice between a ban and a warning label to handle a community that frequently features <speech category>, which would you prefer to have?"* The response categories included: (1) *Community ban:* Platforms should ban this community so that no one can participate in it; (2) *Community warning label:* Platforms should place a warning label before this community but let users who want to participate access it; and (3) *Neither:* Platforms should neither ban this community nor put a warning label on it.

### 3.1.4 Control Variables

I controlled for gender, race, age, education, political affiliation (1 = "very liberal," 7 = "very conservative"), and social media use of each respondent in my regression models. To assess the frequency of social media use, I followed Ernala et al. [27]'s recommendations and asked participants to respond to the question, *"In the past week, on average, approximately how much time PER DAY have you spent actively using any social media sites like Facebook and Reddit?"*

### 3.2 Data Analysis

I began by examining the descriptive statistics of key measures listed above. Next, as mentioned before, I used hierarchical linear regression models to analyze my survey data. I used these models for their ease of interpretability after checking for the underlying assumptions. I created three separate models to evaluate participants' support for platforms' banning of communities featuring hate speech, violent content, and sexually explicit content, respectively. Additionally, I built another three models for evaluating participants' support for platforms putting warning labels before communities featuring each inappropriate speech category.

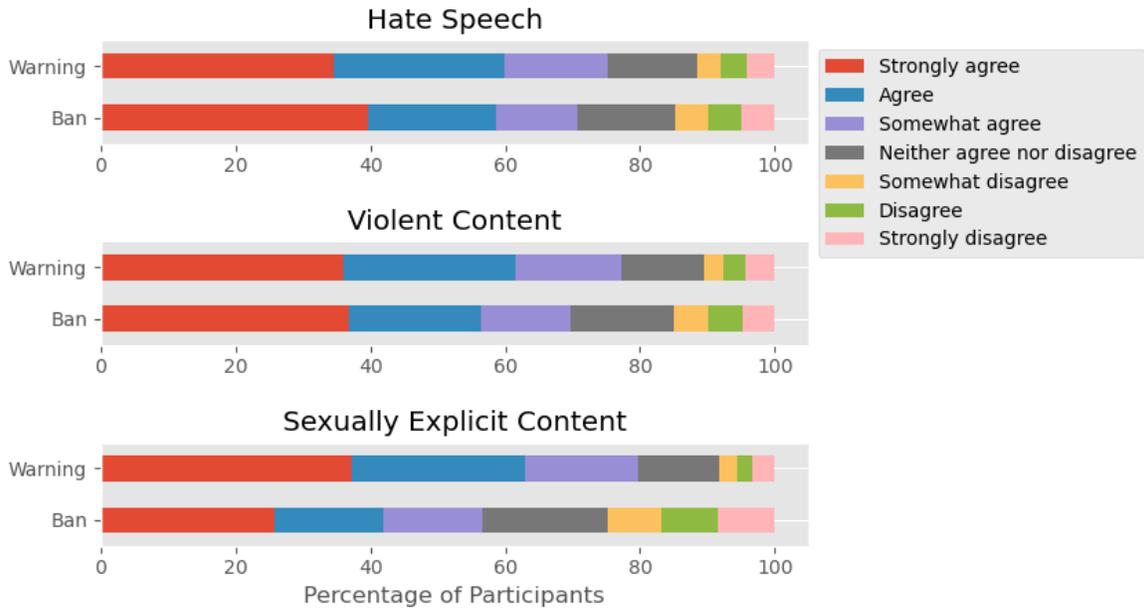

**Figure 3:** Frequency of participants' responses to survey questions about support for bans and putting warning labels before communities featuring hate speech, violent content, and sexually explicit content, measured in percentage.

For analyzing responses to the open-ended question, I used an inductive analysis approach [114] to iteratively develop a set of codes. My research assistant and I coded responses to this question side-by-side to first generate a set of initial codes and describe them. We used NVivo 14, a qualitative data analysis software, to perform this coding. As the analysis matured, we refined these codes further and defined them more sharply. In total, 434 of our open-ended responses were substantive enough to warrant coding; the rest were either blank, incomprehensible to either coder, or did not elaborate on participants' reasoning behind their choices, e.g., one response just mentioned, "community warning label", so we chose not to code it. Some responses were so detailed and multifaceted that we attached multiple codes to them. We double-checked our codes for each response and resolved disagreements through discussions. Next, we conducted a comparison of codes and their associated data with one another. These comparisons allowed us to combine and distill our codes into six key themes that I present as our findings. These themes are shown in **Table 5** and elaborated upon in the Results section.

## 4 RESULTS

I began my analysis by measuring the overall level of support for the two community-wide moderation interventions I tested – bans and warning labels. I found that 70.8%, 69.8%, and 56.6% of participants at least somewhat agreed that platforms should ban communities frequently featuring hate speech, violent content, and sexually explicit content, respectively. Further, 75.2%, 77.3%, and 79.7% of participants at least somewhat agreed that platforms should offer warning labels before communities that frequently feature hate speech, violent content, and sexually explicit content, respectively (see **Figure 3**).

When making a choice between bans, warning labels, or neither to regulate online communities, my analysis shows different trends for different content types (see **Figure 4**). More participants preferred bans (45.1%) over warning labels (44.1%) for handling communities frequently containing hate speech. In contrast, more participants preferred warning labels over bans for regulating communities often featuring violent content (51.1% vs. 39.8%, respectively) and sexually explicit content (61.8% vs. 28.8%, respectively). Thus, *H1*, which predicted that for each type of norm-violating community, the average support for adding warning labels would exceed the average support for banning it, **was only partially supported.**

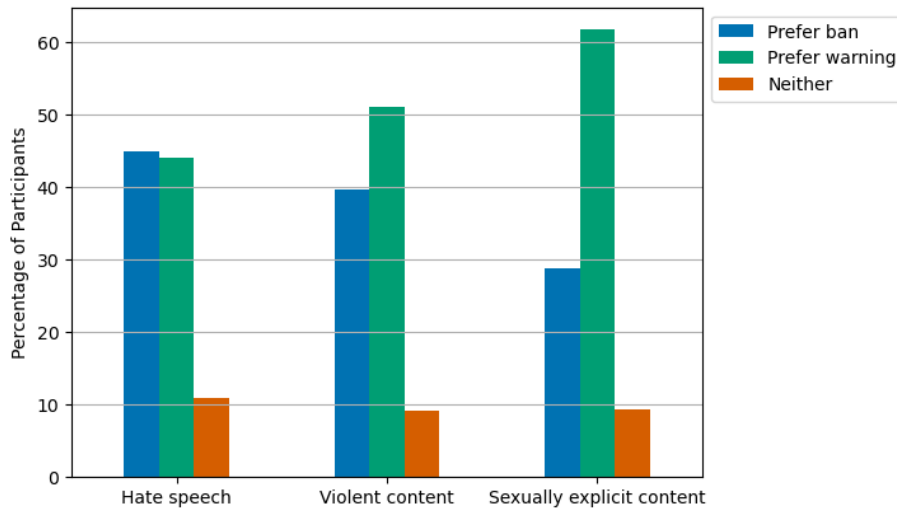

**Figure 4:** Participants' responses to a choice between bans, warning labels, or neither to regulate online communities featuring inappropriate content.

*H2* predicted that for each norm-violating speech category, participants would perceive the effects of communities frequently featuring that content to be stronger on others than on themselves. Paired *t*-tests to assess this found the perceived effects on others to be significantly stronger than on oneself for each category (see **Table 2**). Thus, **my results support *H2*.**

**Table 2:** Descriptive statistics of participants' perceived effects of communities featuring hate speech, violent content, and sexually explicit content on others and self. This table also shows the t-test results comparing perceived effects on others and self (N = 1,023). *** denotes *p* < .001

|  | Effects on | *Mean* | *St. Dev* | *St. Error* | *t* | Cohen's *d* |
|---|---|---|---|---|---|---|
| **Hate speech** | Others | 5.00 | 1.49 | .05 | 15.03*** | .470 |
|  | Self | 4.26 | 1.89 | .06 |  |  |
| **Violent content** | Others | 5.09 | 1.39 | .04 | 15.78*** | .493 |
| | Self | 4.45 | 1.61 | .05 | | |
| **Sexually explicit content** | Others | 4.52 | 1.58 | .05 | 14.66*** | .458 |
| | Self | 3.89 | 1.83 | .06 | | |

## 4.1 Support for Community Bans

I computed hierarchical linear regression to test my hypothesis **H3** and answer **RQ1**. For each norm-violating speech category, I created a model in which the dependent variable was the participants' support for platforms banning communities featuring that category. In Step 1, I included the control variables age, gender, education, race, political affiliation, and social media use. In Step 2, I introduced the independent variables PME3 (perceived effects on others) for that category and support for free speech (**Table 3**).

Table 3: Hierarchical multiple regression analyses predicting support for platforms' banning of communities featuring hate speech, violent content, and sexually explicit content (N = 1,018).

| Independent Variable | Support for platform ban of hate speech ($\beta$) | Support for platform ban of violent content ($\beta$) | Support for platform ban of sexually explicit content ($\beta$) |
|---|---|---|---|
| Model # | Model 1 | Model 2 | Model 3 |
| Step 1 | | | |
| Age | .078** | .069* | .122*** |
| Gender (Female) | .076*** | .138*** | .139*** |
| Race (White) | .030 | .041 | .000 |
| Education[a] | .013 | .028 | -.20 |
| Political affiliation[b] | -.148*** | -.118*** | .010 |
| Social media use[c] | .047 | .044 | .052 |
| $R^2$ | .062*** | .079*** | .068*** |
| Step 2 | | | |
| Support for free speech | -.054 | -.049 | -.025 |
| Perceived effects of hate speech on others | .404*** | - | - |
| Perceived effects of violent content on others | - | .446*** | - |
| Perceived effects of sexually explicit content on others | - | - | .473*** |
| $R^2$ change | .156*** | .187*** | .215*** |
| Total $R^2$ | .218 | .266 | .283 |

*p < .05, **p < .01, ***p < .001 (t test for $\beta$, two-tailed; F test for $R^2$, two-tailed).
[a] 0= Less than secondary education; 1= Secondary education or more.
[b] 1= Strong Democrat, 7= Strong Republican.
[c] 1= Less than 10 minutes per day, 6= More than 3 hours per day.
$\beta$ = Standardized beta from the full model (final beta controlling for all variables in the model).

For each norm-violating speech category, the regression models show significant influences of the participants' perceived effects of communities featuring that category on others (PME3) on their support for a platform ban of such communities (Model 1: hate speech − $\beta$ = .404, p < .001; Model 2: violent content − $\beta$ = .446, p < .001; Model 3: sexually explicit content − $\beta$ = .473, p < .001), **supporting H3.**

On the other hand, greater support for free speech *does not* influence support for platform bans of communities in *any* category: hate speech ($\beta$ = -.054, $p$ > .05), violent content ($\beta$ = -.049, $p$ > .05) or sexually explicit content ($\beta$ = -.025, $p$ > .05). This **answers RQ1**.

## 4.2 Support for Community Warning Labels

I computed hierarchical linear regression to test hypotheses **H4** and **H5**. For each norm-violating speech category, I created a model where the dependent variable was participants' support for putting warning labels before communities frequently containing that content category. Similar to models for community bans, in Step 1 of the three regression models, I included the control variables age, gender, education, race, political affiliation, and social media use. In Step 2, I introduced the independent variables PME3 (perceived effects on others) for that category and support for free speech (**Table 4**).

**Table 4:** Hierarchical multiple regression analyses predicting support for platforms putting warning labels before communities frequently featuring hate speech, violent content, and sexually explicit content (N =1,018)

| Independent Variable | Support for warning labels before hate speech ($\beta$) | Support for warning labels before violent content ($\beta$) | Support for warning labels before sexually explicit content ($\beta$) |
|---|---|---|---|
| Model # | Model 4 | Model 5 | Model 6 |
| Step 1 | | | |
| Age | .052 | .007 | .045 |
| Gender (Female) | .072* | .036 | .072* |
| Race (White) | .016 | .020 | -.026 |
| Education[a] | -.024 | .038 | -.038 |
| Political affiliation[b] | -.025 | -.010 | -.093** |
| Social media use[c] | .045 | .083** | .075* |
| R$^2$ | .020** | .018** | .023*** |
| Step 2 | | | |
| Support for free speech | .212*** | .242*** | .237*** |
| Perceived effects of hate speech on others | .242*** | - | - |
| Perceived effects of violent content on others | - | .220*** | - |
| Perceived effects of sexually explicit content on others | - | - | .220*** |
| R$^2$ change | .112*** | .119*** | .108*** |
| Total R$^2$ | .132 | .137 | .131 |

**p < .01, ***p < .001 (*t* test for $\beta$, two-tailed; F test for R$^2$, two-tailed).
[a] 0= Less than secondary education; 1= Secondary education or more.
[b] 1= Strong Democrat, 7= Strong Republican.
[c] 1= Less than 10 minutes per day, 6= More than 3 hours per day.
$\beta$ = Standardized beta from the full model (final beta controlling for all variables in the model).

For each norm-violating speech category, the regression models show significant influences of the participants' perceived effects of communities featuring that category on others (PME3) on their support for using warning labels before those communities (Model 4: hate speech − $\beta$ = .242, $p$ < .001; Model 5: violent content − $\beta$ = .220, $p$ < .001; Model 6: sexually explicit content − $\beta$ = .220, $p$ < .001), **supporting H4**.

Greater support for free speech has a significant positive influence on participants' support for using warning labels before communities featuring each norm-violating category (Model 4: hate speech – $\beta$ = .212, $p$ < .001; Model 5: violent content – $\beta$ = .242, $p$ < .001; Model 6: sexually explicit content – $\beta$ = .237, $p$ < .001), **supporting *H5*.**

Other notable insights from my analyses relate to how demographic characteristics and social media use affect bystander perceptions. Age was positively related to support for community-wide bans for each speech category. Females showed greater support for community-wide bans in all speech categories compared to males. Democrats were more likely than Republicans to support community-wide bans of hate speech and violent content, but not sexually explicit content; however, Democrats were more likely than Republicans to support warning labels before communities featuring sexually explicit content. The frequency of social media use was positively related to support for warning labels before communities featuring violent and sexually explicit content, but not hate speech.

### 4.3   Qualitative Analysis of Open-Ended Responses.

I report below the results of the qualitative analysis of open-ended responses to the survey question that asked participants to reflect on the reasoning behind their community moderation choices. These results add more nuance to the findings presented above. **Table 5** summarizes the themes describing perspectives that influenced or explained participants' community moderation choices.

**Table 5:** Perspectives that influenced content moderation choices selected by participants. Frequency shows the number of respondents who displayed each perspective (Total = 434).

| Theme | Frequency |
|---|---|
| Desire for freedom of choice in community engagement | 143 |
| Norm-violating communities could reinforce inappropriate behavior | 82 |
| Effectiveness of sanctions in reducing inappropriate speech | 42 |
| Need for nuance in enacting content moderation | 79 |
| Impact of speech on others | 47 |
| Ethical objections against content | 95 |

Next, I describe each theme by highlighting excerpts of respondent quotes that serve as representative examples.

*4.3.1 Desire for Freedom of Choice in Community Engagement*

143 participants (32.9%) noted that their inclination to have a choice in whether to engage with community content influenced their choice of sanction for the norm-violating communities they encountered during the survey. Many of these participants expressed strong support for freedom of expression and offered it as justification for their opposition to community bans (although my quantitative findings regarding *RQ1* did not detect a significant influence between these two measures). For example, Participant P157 argued that "an outright ban goes against the morals and values of democracy." Similarly, Participant P1006 wrote:

> *Freedom of Speech is a foundational cornerstone in the construct of the Western World's strongest, most powerful and world-influential government. Any attempt to ban, restrict, prohibit or police any form of communication or the expression of ideas, philosophies, ideologies or beliefs is contrary to that government's founding principles.*

Most of these participants selected to insert a warning label before norm-violating communities, arguing that the use of such labels preserves freedom of expression and empowers users to decide for themselves whether they want to engage with the labeled community. This aligns with my quantitative findings in support of **H5**. For example, Participant P338 wrote,

> *"Generally, I support a warning label rather than a full ban; a warning label lets people choose with the knowledge of what they are getting into instead of making the decision of what content is available to them over their heads."*

### 4.3.2 Norm-violating Communities Could Reinforce Inappropriate Behaviors

82 participants (18.9%) expressed concern that norm-violating communities, if they remained unregulated, would encourage users who engage with community content to behave inappropriately, either online or offline. Therefore, they preferred that platforms ban such communities. They felt that joining such communities would facilitate repeated exposure to hate speech, violent content, or sexually explicit content and normalize such behaviors. Participant P812 pointed out, "We are all influenced by the things we see and hear and a constant spread of this influences what we do." These respondents worried that norm-violating communities could offer a sense of belonging to alienated individuals and allow them to become more extremist in their views. For example, Participant P840 noted that membership in such communities could "encourage herd-mentality and whip up people to more and more hate, violence, etc." Similarly, Participant P275 wrote:

> *"Anytime a platform urges to hate or discriminate and can influence others is extremely dangerous."*

16 participants were especially concerned that violent communities pose risks to public safety because they could influence their members to engage in acts of violence. Participant P778 hoped that banning such communities "would hopefully curb some of the mass shootings that have been happening in our country." Similarly, Participant P758 noted:

> *"Violence of any kind or hate towards other humans can lead people that may already be unstable to commit violent acts against other people or certain groups of people."*

41 of these participants were worried about children's exposure to inappropriate online communities. They felt children are especially vulnerable to embracing the values spread by such communities and acting upon them. Participant P716 argued:

> *"I feel that violence on any platform is not good for anyone and could possibly be accessed by children, shaping them and their developing minds in a very negative way."*

These findings add more nuance to my quantitative results supporting **H3** and **H4**.

### 4.3.3 Effectiveness of Sanctions in Reducing Inappropriate Speech

42 respondents (9.7%) mentioned that their choice of sanctions was influenced by their perceptions of the effectiveness (or *ineffectiveness*) of those sanctions in reducing the spread of norm-violating content in the future. 18 of these participants preferred community bans for all three norm-violating content categories in the survey and justified this selection by arguing for their effects on future postings. For example, Participant P1003 supported community bans by contending that "community bans cut the problem at its root." Similarly, Participant P803 wrote,

> *"Community ban would make sure these things would be stopped because if you ban the community, it can't happen anymore."*

Along these lines, 10 participants argued for the use of warning labels, citing their perceived effectiveness in curbing norm-violating speech. These participants felt that putting up a warning label would alert the community in question and potentially trigger a self-correction by community members and moderators. Both P305 and P318 insisted that everyone should have at least one chance to correct themselves. Similarly, Participant P351 noted,

> *"I think people should learn from their actions instead of immediately being banned."*

In contrast, 10 of these 42 respondents were not convinced of the effectiveness of community bans in reducing the spread of inappropriate content. They felt that shutting down inappropriate speech through moderation does not stop the ideas behind it because people always find a way to express their views. For example, Participant P823 pointed out,

> *"If we ban communities entirely it could cause them to voice their reasons in real life which could lead to violence and targeted attacks."*

Similarly, 8 participants argued against the use of warning labels by noting their perceived ineffectiveness in reducing the spread of inappropriate speech. For example, Participant P503 maintained that "a warning is like a slap on the wrist that allows people to go back and do it over and over again." Participant P1004 wrote,

> *"What is a warning label going to do? Might even make someone more interested [in the labeled community]."*

### 4.3.4 Need for Nuance in Enacting Content Moderation

79 participants (18.2%) argued for platforms to enact content moderation with more nuance and on a case-by-case basis. They asserted that community-wide bans can be too coarse and emphasized the importance of meting out sanctions in proportion to the severity and type of norm violations. For example, Participant P236 wrote:

> *"As far as violence, I think it depends on the context...A full-on ban on violent content wouldn't really be a great idea because there is content that might be considered violent but isn't harmful in any way, such as content from TV/movies or video games, etc. There is also some real-life violent content that I think deserves to have a place on the internet, for example, bodycam footage from police."*

Similarly, Participant P1042 noted:

> *"It depends on the specifics. For example, if someone is calling for harm against a group or person and hurts someone on screen, that should probably be banned. But in many cases, violent content sheds light on violent scenarios such as genocides that benefit from awareness."*

23 of these respondents insisted that the concepts of hate speech or violent speech are too broad and subjective to warrant a sanction as severe as community-wide bans. For example, Participant P768 wrote that "what is offensive to one person may not be offensive to another... as soon as you restrict one person, group or idea – in principle what stops you from restricting another?" Participant P969 noted, "My concern is how to define hate speech, and if you block it ... haters will find another way to spew hate." Participant P335 pointed out:

> *"Cancel culture can go too far both ways and there needs to be more clarification on some of these topics like violence and hate speech."*

### 4.3.5 Impact of Speech on Others

47 participants (10.8%) mentioned that their sanction choices (or lack thereof) were influenced by their concerns about the consequences of inappropriate speech on others. These respondents often emphasized that they are accepting of controversial speech so long as it does not harm others. For instance, Participant P422 wrote, "Say whatever as long as it doesn't affect anyone else." Participant P879 preferred that sexually explicit communities not be banned because in his view, "sexual content is, generally, not as damaging or dangerous when exposed to, even if exposure occurs accidentally or fraudulently."

25 of these participants explicitly mentioned that when an online community inflicts harm such as harassment, hate, or violence against others, especially those belonging to marginalized or vulnerable groups, platforms should sanction that community. For example, Participant P530 worried about situations in which a single individual may experience a coordinated attack by a community; she preferred in such cases that the community be banned to protect victims' mental health. Most of these participants preferred to ban communities featuring hate speech or violent content, citing their potential to inflict physical or emotional harm on others. For example, Participant P338 wrote:

> *"In the case of hate speech, certain individuals or groups may be ENDANGERED by the speech if it includes threats or the discussion of plans to do harm. In such instances, there should be a ban to prevent the commission of crimes against targeted individuals and groups, and perhaps legal action should be taken."*

*4.3.6 Ethical Objections Against Content*

95 participants (21.9%) expressed their ethical objections against hate speech, violent content, or sexually explicit content to justify their choice of sanctions against communities featuring them. For example, Participant P956 asserted that "hate, violence, and stereotyping should not be normalized or desensitized." Similarly, Participant P1020 pointed out:

> *"I think racism should be banned. Hate speech like that is not okay in any circumstance and should not be spread."*

18 of these participants further argued that social media platforms hold a responsibility to mitigate the harmful influence of ethically inappropriate content. For example, Participant P782 noted that social media "platforms have a duty to the society we live in" and "there needs to be a moral compass" dictating their regulation. Participant P984 felt:

> *"These social media platforms cannot be a safe haven for hate speech, violence, or sexually explicit content. They've reaped the financial benefits of these platforms but have til (sic) now shown no responsibility for allowing the content to spread like wildfire when it's been shown over and over these platforms add to the extremism of the country."*

7 participants supported their selection of norm-violating community bans by explicitly rooting their argument in their religious ethics. For instance, Participant P615 pointed out that he prefers to use platforms that support his Judeo-Christian values. Similarly, Participant P995 wrote,

> *"I am a Christian and I don't support hate, violence, and public sexual acts."*

## 5 DISCUSSION

Prior research on community-wide moderation interventions has examined their effects on community members' activities on the platform (either within the sanctioned community or elsewhere on the platform) and on other websites that they migrate to [14, 15, 47, 100, 120]. The current article builds upon this rich literature to understand bystanders' reactions to such interventions. It is not immediately apparent how bystanders would perceive community-wide sanctions. On the one hand, they do not target a specific individual, and this absence of calling someone out as a violator may reduce personal stake in the moderation decision and its effects. On the other hand, sanctions that simultaneously impact an entire community can be perceived as overly broad, as they also affect users whose contributions to the said community did not violate platform rules.

### 5.1 Support for Community Bans versus Community Warning Labels

This study shows that most US adults at least somewhat support both moderation interventions (i.e., bans and warning labels) for communities featuring each norm-violating content category I tested. For many users, these preferences are grounded in their core ethical and religious values. Given a choice between a ban, a warning label, or neither intervention, only about 10% of participants opted to select neither intervention in each case. *This widespread acceptance of community-wide moderation interventions should empower platforms to take such actions when community content shows clear, persistent patterns of norm violations.* My qualitative analysis also shows that many users' attitudes toward moderation sanctions are shaped by their perceived empirical effectiveness in reducing future norm violations. Thus, platforms could publicly share reports that document the statistical effects of prominent sanctions on various metrics of site activity. These findings also contribute to ongoing conversations about how to negotiate power dynamics between different levels of platform governance [56], e.g., when should platform administrators intervene in the governance of an online community rather than let it self-govern its activities.

The support levels for selecting bans versus warning labels vary across the three content categories, with more participants preferring to ban communities frequently featuring hate speech rather than putting warning labels on them. However, participants preferred warning labels over bans for communities with violent and sexually explicit content. This finding adds more nuance to prior literature, which observed a preference for warning labels over bans for handling misinformation [5, 127]. It shows that users prefer moderation approaches of varying severity for different categories of norm-violating content. This insight is further supported by the survey's open-ended responses desiring a greater nuance in how different norm-violating content categories are defined, interpreted, and sanctioned. Therefore, *platforms should characterize community-wide norm violations of speech content more granularly and deploy sanctions proportionate to the offense.*

Prior HCI research has characterized content removal as a punitive measure and acknowledged its inherent limitations [33, 130]. To address these limitations, researchers have recommended that platforms establish publicly visible content curation guidelines, notify posters about their post removals, and offer explanations for content removals [50, 53, 75]. Alternative justice frameworks such as restorative justice and transformative justice have also been proposed as frameworks that could repair online harm and restore individuals and communities [81, 105, 129, 130]. Given the widespread support for inserting warning labels before norm-violating communities uncovered in my survey data, I advocate for them as another educational strategy to shape users' understanding of accepted platform norms and motivate them to improve future behavior on the site [126].

## 5.2 Influence of Third-Person Effects on Support for Community-wide Moderation

Study findings support the perceptual component of the TPE hypothesis, revealing that participants believed communities featuring hate speech, violent content, and sexually explicit material had a greater influence on others than on themselves. Additionally, for each content category, participants' perception of the effects on others (PME3) was a significant predictor of their support for both community-wide bans and warning labels. These findings suggest that when users perceive online communities as detrimental to the broader user population, they desire social media platforms to take comprehensive actions against those communities. My qualitative analysis supports this interpretation by showing that the perceived influence and damaging impact of speech on others strongly motivates many users' desire for platform-enacted sanctions. Further, I found that many users worry that norm-violating content could influence suggestible others, especially children, to become more extremist. Community-wide moderation actions often generate broad discontent and discussions about their merits among platform users in the aftermath [14, 15, 47]. My insights about PME3 suggest that *platforms could bolster support for such actions by publicly detailing how content posted on the sanctioned communities poses risks to vulnerable others.*

## 5.3 Influence of Support for Free-Speech on Support for Community-wide Moderation

I found *no evidence of any relationship between free speech support and community-wide ban support for each of the three norm-violating categories*. This finding shows that individuals' free speech values do not automatically make them more accepting of harmful online communities. This result is also consistent with the mixed findings for the relationship between free speech support and supportive attitudes toward platform censorship observed in prior literature [43, 58]. One possible explanation for this result is that my participants perceived these three norm-violating content types as the sort of expressions that fall outside the ambit of free speech protections [48]. Another contributing factor could be a broader public awareness about the exploitation of free speech values to permit the spread of inappropriate content [30, 125, 128], especially because of the recent emergence of Alt-tech platforms [133] that market themselves as free-speech havens but attract conspiracy theorists and trolls.

On the other hand, I found that *support for free speech predicted support for warning labels before communities featuring each inappropriate speech category.* This suggests that people may perceive warning labels not as a violation of others' freedom of speech but rather as a means to empower themselves and others in choosing the content they consume. Participants' open-ended responses also show that many free-speech advocates welcomed having a choice in whether to engage with the community content via informative prompts in warning labels. These insights add further evidence to prior scholarship suggesting that for many users, free speech principles can comfortably align with certain content moderation practices [58, 96]. Free speech scholars can expand on this work by examining how individuals' support for freedom of expression influences their perceptions of the communities they are active in being labeled under a norm-violating category.

Although not the main focus of this study, my analyses show how age, gender, political affiliation, and social media use are associated with respondents' content moderation preferences. Future research on *why* specific demographic groups respond to online harms differently could offer valuable insights into building content curation policies and moderation resources that sufficiently address users' diverse needs.

### 5.4 Limitations and Future Work

This study focuses on how detached bystanders would perceive sanctions of norm-violating online communities. It would be interesting to explore how community members themselves perceive such actions. While prior research has examined changes in the posting activity of members of sanctioned communities [14], understanding their views on such sanctions could further inform platform decisions.

Second, my survey questions were not grounded in a specific platform to ensure broader applicability. Future studies focusing on particular social media sites could explore whether users' attitudes toward specific platforms influence their perceptions of community-wide moderation actions on those platforms.

Third, I sought respondents' opinions on communities featuring hate speech, violent content, and sexually explicit content since platforms emphasize them in their guidelines and reporting interfaces [134]. However, these content categories could be perceived as overly broad. For example, while most people would agree that content featuring nonconsensual pornography, sex trafficking, and sexual exploitation of minors deserve regulation, other sexually explicit posts featuring artistic representations of the human body or sex education would likely be deemed appropriate. Thus, future work must look beyond the reductive ways in which platforms conceptualize looking at these content categories. Relatedly, the survey items for measuring the impacts of sexually explicit content used in this study could be improved to better accommodate individuals with LGBTQ+ identities.

Fourth, while the regression models used in this study offer valuable insights into the relations between various factors and moderation preferences, they do not allow for making causal claims. It is possible that these models did not account for relevant mediating factors, such as participants' own previous postings of inappropriate content and prior moderation sanctions against them or their perceptions of social norms about community-wide moderation. However, my analysis of participants' open-ended responses largely aligns with my quantitative findings and adds confidence to my modeling. Future work can improve upon this study design by conducting online social (pseudo)experiments that rule out potential confounds.

Finally, my survey questionnaire asked respondents to choose sanctions for communities based on just their problematic norm violations. However, users' perceptions of any community sanctions would be influenced not just by their violations but also by the potential value those communities offer. Therefore, future research could test bystander attitudes about specific instances of online community sanctions by showing respondents more details about the community, such as a sample of posts that appear on it.

## 6 CONCLUSION

This paper examines social media users' attitudes toward community-wide moderation sanctions and the factors that shape those attitudes. My findings highlight that users' moderation preferences are strongly shaped not just by their own ethical objections against norm-violating content but also by their perceptions of how other users engage with or could be affected by such content. Further, I found that users' free speech values comfortably align with their preference for adding warning labels to inappropriate content. The results of this study begin to offer fundamental insights into Americans' perceptions of what platforms are, how they assert their power in society, and what role they should play in shaping public discourse. I call for more user-centered research to further the development of platforms in ways that align with the general public's values and address their safety needs.

## ACKNOWLEDGMENTS

Acknowledgments will be added after the peer review is completed.